\documentclass[10pt,a4paper]{article}

\def\heading #1{\bigbreak \begin{center} {\bf #1} \end{center}}

\title{Annihilation rate in positronic systems by quantum Monte Carlo. e$^+$LiH as test case.}

\author{Massimo Mella$^a$ and Simone Chiesa$^b$\\
Dipartimento di Chimica Fisica ed Elettrochimica,\\ Universit\'a degli Studi
di Milano, via Golgi 19, 20133 Milano, Italy\\
$^a$Electronic mail: Massimo.Mella@unimi.it\\
$^b$Electronic mail: Simone.Chiesa@unimi.it\\
\and 
Gabriele Morosi$^c$\\
Dipartimento di Scienze Chimiche, Fisiche e Matematiche,\\
Universit\'a dell'Insubria, \\
via Lucini 3, 22100 Como, Italy\\
$^c$Electronic mail: Gabriele.Morosi@uninsubria.it \\}

\begin{document}

\maketitle

\bigbreak
\begin{abstract}

An accurate method to compute the annihilation rate in positronic systems 
by means of quantum Monte Carlo simulations is tested and compared 
with previously proposed methods using simple model systems. This method can be
applied within all the quantum Monte Carlo techniques, 
just requiring to accumulate the positron-electron distribution
function. The annihilation rate of e$^+$LiH as a function of the internuclear 
distance is studied using a model potential approach to eliminate the core 
electrons of Li, and explicitly
correlated wave functions to deal with all the remaining particles.
These results allow us to compute vibrationally averaged annihilation rates,
and to understand the effect of the Li$^+$ electric field on positron and
electron distributions.

\end{abstract}

\subsubsection*{PACS number(s): 36.10.-k, 02.70.Lq}

\pagebreak

\heading{I. INTRODUCTION}

In positron and positronium (Ps) chemistry and physics,
the annihilation rate $\Gamma_{2\gamma}$ plays an important role
since it correlates with many aspects of the local environment where
the positron annihilates. 
For instance, ``pick-off'' annihilation in solutions and in solid
materials, ``on the fly'' annihilation in atomic and molecular gases, and
bound state annihilation of positronic compounds are just few of
the experiments where $\Gamma_{2\gamma}$ can be measured and 
successively interpreted.

Although these experiments are relevant both technologically and 
scientifically ~\cite{moge,kraus}, only few theoretical studies have been 
devoted to accurately compute
annihilation rates for realistic systems like atoms and molecules
in order to compare with experimental data or to predict trends
~\cite{mitroy,newjim,jiang_psh,stras3,stras4,mecor,morosi,maxpsh,maxcor}.
Moreover, these have been restricted to deal at most with four active electrons,
so that only a bunch of systems have been studied so far.
We believe this scarceness of results to be primarily due to the intrinsic 
difficulty in obtaining accurate wave functions for larger systems, and to the
computational effort requested with respect to ordinary matter compounds 
when standard {\it ab initio} methods are employed ~\cite{brom1}.

For these reasons, quantum Monte Carlo (QMC) methods ~\cite{reybook} 
represent an alluring alternative to these methods, to Density Functional 
Theory,
and to explicitly correlated wave functions in computing energies and properties
of realistic positronic systems. 
QMC techniques are well described in the literature, so we avoid to burden
this paper with the details of the methods and constrain ourselves to discuss
only the technical issues relevant for the specific problem.

Not requiring the analytical calculation
of integrals, QMC allows one to use any physically sensible wave 
function. This possibility increases the chances to obtain an accurate 
description of any class of systems once all the relevant physical 
information is included in the chosen analytical form of the wave function. 
Having defined a trail wave function $\Psi_T$ for a system, 
QMC techniques allow one to compute the differential and 
non-differential properties of the system by sampling $\Psi_T^2$, 
$\Psi_T\Psi_0$, or $\Psi_0^2$. Here, $\Psi_0$ is the exact ground state
function of the system.
This task is usually accomplished by creating a distribution of points 
(also known as configurations or walkers)
in configurational space whose density is proportional to the aforementioned
$\Psi_T^2$, $\Psi_T\Psi_0$, or $\Psi_0^2$.

Keeping in mind the above remarks, it might appear that the QMC methods should 
accurately predict any interesting observable for positronic systems.
This is indeed correct except for extremely local operators like Dirac's 
delta $\delta$, and hence for $\Gamma_{2\gamma}$ that is proportional to
its expectation value, for which an accurate sampling of small configurational 
space volumes is needed. These operators are well known to represent a 
challenge for QMC 
due to the discrete nature of the configuration ensemble and the finite length
of the simulations.

As far as the mean value of the Dirac's delta $\delta$ operator is concerned,
one faces an additional difficulty when trying to estimate its mean value. 
Even admitting a perfect sampling
in the regions where two particles are close to each other, the primitive 
method of counting the number of times
the interparticle distance $r$ is smaller than a given radius $r_w$ 
(i.e. counting the ones that fall into a spherical well of radius $r_w$) 
~\cite{jiang_psh} has an 
associated statistical error that diverges for $r_w\rightarrow 0$
~\cite{maxcor}. 
This fact means that the estimation of the statistical
error of the extrapolated value is based on shaky grounds. 

Although not a solution, a slightly better approach was devised by 
substituting the simple sphere with a Gaussian function centered
at the coalescence point ~\cite{mecor,morosi,maxcor}. 
The variance of this estimator also goes to infinity
upon decreasing of the Gaussian width, but it diverges less faster than the 
one of the spherical well, therefore allowing a
statistically more accurate estimation of $\langle \delta (r_{-+}) \rangle
=\sum_i \langle \delta (r_{i+}) \rangle$.

Due to the interest in computing $\langle \delta (r) \rangle$ for many systems,
attention has been paid to solve these problems,
and remedies have been suggested in the framework of all the QMC techniques.

As far as variational Monte Carlo (VMC) is concerned, different
methods have been proposed that may solve this difficulty allowing one 
to compute the needed quantities. One of these methods starts from the 
distribution differential identity

\begin{equation}
\nabla^2 \frac{1}{r} = -4\pi \delta(r)
\label{eq1}
\end{equation}
\noindent
that allows one to write, after specializing for the electron-positron pair
~\cite{maxpsh}

\begin{eqnarray}
\label{eq2}
\langle \delta(r_{1+}) \rangle_T =
\int \delta(r_{1+}) \Psi_T^2(\mathbf{R}) d \mathbf{R} =\\ \nonumber
-\frac{1}{2 \pi} \int \Psi_{T} ^2 ({\mathbf R}) \{
\frac{\nabla ^2 _{\mathbf{r}_{1}}\Psi_{T}({\mathbf R})}{\Psi_{T}({\mathbf R})}
+ [ \nabla _{\mathbf{r}_{1}}\ln \Psi_{T}({\mathbf R}) ]^2
\}\frac{1}{r_{1+}} d{\mathbf R}
\end{eqnarray}
\noindent 
where $\mathbf{R}=(\mathbf{r}_1,\mathbf{r}_2,...,\mathbf{r}_+)$ 
is a point in configuration space, and the trial wave function $\Psi_T$ 
is normalized.
Although this integral has a well defined value that can be computed
sampling $\Psi_T^2$, it is well known that its variance diverges
over the same distribution ~\cite{langf}. 
This fact implies that no error bound (i.e.
standard deviation) can be associated to its value, a dangerous
situation one would like to avoid.

Langfelder {\it et al.} ~\cite{langf} proposed a possible way to
circumvent this problem
based on a modified importance sampling transformation where
$\Psi_T^2 \sum_i 1/r^2_{i+}$ is sampled instead of $\Psi_T^2$. 

Always starting from Eq. \ref{eq2}, one could also exploit the approach
proposed by Assaraf and Caffarel ~\cite{caffarel} to compute the expectation 
value needed to obtain nuclear forces by means of the Hellman-Feynmann theorem.
They showed that a judicious choice of a renormalized operator, whose
mean value is equal to the original one, can reduce 
the infinite variance to a finite value ~\cite{caffarel2}.

A completely different approach was pursued by Alexander and Coldwell
~\cite{coldw}. They
proposed to compute all the mean values sampling an analytically normalizable
distribution function $g(\mathbf{R})$, so that the normalization 
$N_T$ of $\Psi_T$ is easily estimated by means of 

\begin{equation}
\frac{1}{N_T^2} = M^{-1} \sum_{i=1}^M \Psi_T^2(\mathbf{R}_i)/g(\mathbf{R}_i)
\label{eq3}
\end{equation}
\noindent
where the $M$ points sample the normalized $g$. If a second 
normalizable distribution $g_c(\mathbf{R})$, constrained on the subspace 
$\mathbf{r}_+ = \mathbf{r}_1$, is employed to guide the simulation and 
to compute $N_T^c$ in Eq. \ref{eq3}, then 
$\langle \delta(r_{-+}) \rangle$ can be easily estimated by the 
$(N_T^c/N_T)^2$ ratio.

Although these three methods represent a step towards the solution of this
complicate problem in the VMC framework and are currently used for ordinary electronic compounds with success, the situation still remains 
far from being satisfactory for positronic systems.
For these systems beyond the problem of the method used to compute 
$\langle \delta \rangle$, there is another difficulty:
as far as we know, nobody has been able to optimize 
an accurate $\Psi_T$ for a positronic system with more than four electrons. 
More specifically, for large systems
explicit correlation between the electrons and the positron has been found
difficult to introduce ~\cite{maxcor}.
This means that the "pile up" of the electron density
over the positron is not correctly described, 
therefore giving rise to too small annihilation rates ~\cite{spiega1}.
Possible sources of this unwanted outcome are the lack of knowledge about
the complicate analytical form that such an accurate wave function should
have, and some drawbacks of the optimization method used ~\cite{spiega2}.

In order to go beyond these difficulties, the diffusion Monte Carlo (DMC) method
is usually employed to sample $\Psi_T \Psi_0$ ~\cite{and75,and76,rey82}.
This technique is able to project out the contribution of the 
excited states from the starting $\Psi_T$, allowing the exact calculation of the
ground state energy. Unfortunately, the $\delta(r_{-+})$ operator
does not commute with the Hamiltonian of the system, so that the simulation
results are only an approximation to the exact mean value when
computed by means of the mixed estimator

\begin{equation}
\langle \delta(r_{-+}) \rangle_M =
\int \delta(r_{-+}) \Psi_T \Psi_0 d \mathbf{R}
\label{eq4}
\end{equation}
\noindent
Although, this value represents a more accurate estimate 
of the exact $\langle\delta(r_{-+})\rangle$ than 
$\langle\delta(r_{-+})\rangle_T$, it has been found that the quality of the 
results strongly depends on how accurately $\Psi_T$ mimics the correct 
interparticle distributions.

Whereas both the spherical well and Gaussian method can be employed to
estimate $\langle\delta(r_{-+})\rangle_M$ in Eq. \ref{eq4},
Jiang and Schrader ~\cite{jiang_psh} pointed out that the use of the 
differential identity Eq. \ref{eq1} in a DMC simulation requires some 
uncontrolled
approximation, since $\Psi_T \Psi_0$ is not known analytically but only sampled.

Nevertheless, it has been shown ~\cite{maxpsh} that 
an accurate estimate of $\langle\delta(r_{-+})\rangle$ can be obtained 
simply substituting $\Psi_T^2$ in Eq. \ref{eq2} with $\Psi_T \Psi_0$,
if $\Psi_T$ correctly describes the positron-electron distribution.

A possible solution to the difficulty that DMC meets in estimating
the exact expectation values is represented by sampling $\Psi_0^2$
instead of $\Psi_T \Psi_0$, and computing $\langle \delta(r_{-+}) \rangle$
without resorting to $\Psi_T$ in any way. This idea rules out the possibility
to use Eq. \ref{eq1}, since one just samples the exact $\Psi_0^2$
distribution and no analytical information is available about its form.

In order to overcome this problem, Langfelder {\it et al.} ~\cite{langf}
proposed to correct $\langle \delta(r) \rangle_T$
by accumulating the walker weights in a small sphere around the
coalescence point. Although this may look as a promising way,
we noticed in our work on positron complexes ~\cite{pshpol}
that long decaying times
are needed in order to project out all the excited state contributions
and to correctly build the "pile up" of the electron density over the positron
if $\Psi_T$ poorly describes this feature.
This fact produces large fluctuations in the weight values, therefore
increasing the statistical noise of the results.

A better approach may be represented by the use of the tagging algorithm
proposed by Barnett {\it et al.} ~\cite{reytag} in connection with the 
branching step
usually employed in DMC. Here, the ratio $\Psi_0 /\Psi_T$, needed
to sample $\Psi_0^2$, is computed by means of the number of daughters
of each configuration.

Moreover, Baroni and Moroni ~\cite{rept} have recently proposed an alternative 
algorithm that appears to be well suited for this task. This is based on a
``Path Integral'' view of the DMC algorithm, where the branching step
has been substituted by an accept/reject step in order to exactly sample
$\Psi_0^2$.

Unfortunately, these approaches do not solve the problem of the 
scarce sampling in the volume around $r=0$, a problem that is present 
even for small simulation time steps. As stated previously,
this comes from the finite length and discrete nature of the QMC simulations.
As an attempt to overcome this difficulty, Langfelder {\it et al.} 
~\cite{langf} implemented in their algorithm the 
correct sampling of the electron-nucleus cusp region as proposed by 
Umrigar {\it et al.} ~\cite{umri_dmc}:
this, however, does not appear straightforward to adopt to correct the 
sampling of both the electron-electron and electron-positron cusps.

Keeping in mind all the aforementioned problems in estimating 
$\langle \delta \rangle$, we believe the Monte Carlo  practitioners are left 
only with the hope of devising an approximate, but hopefully 
solid and accurate, method to compute this observable.

The main aim of this paper is to discuss and test the accuracy of computing
$\langle \delta \rangle$ using some simple methods based only on the sampling of
the positron-electron distribution function without any usage of the 
differential identity Eq. \ref{eq1}. These methods will be compared
with the Gaussian approximation discussing relative merits and
applicability. Moreover, we apply them to the realistic e$^+$LiH model case
in order to study the annihilation rate as a function of the internuclear 
distance R. The $\Gamma_{2\gamma}$ versus R results will
allow to compute the vibrationally averaged annihilation rate for this system
and to discuss molecular environment effects on the annihilation rate itself
and on contact distribution functions.

The outlines of this work follow. In Section II we present the
basis of the methods. Section III describes their applications to model systems
for which the exact $\langle \delta \rangle$'s are known. 
As application of
this technique, we deal in Section IV with the model e$^+$LiH.
Our conclusions and proposals of future work are then presented in Section V.

\heading{II. METHODS}

Since we want to develop a method that can be applied to any QMC
technique, henceforth we will use $f(\mathbf{R})$ to indicate cumulatively 
$\Psi_T^2(\mathbf{R})$, $\Psi_T(\mathbf{R})\Psi_0(\mathbf{R})$, or 
$\Psi_0^2(\mathbf{R})$. 
Here, $\mathbf{R}=(\mathbf{r}_1,\mathbf{r}_2,...,\mathbf{r}_+)$ is a point
in configuration space, $\mathbf{r}_i$ and $\mathbf{r}_+$ being 
respectively the i-th electron and positron positions. 

We are interested in computing the expectation value 
$\langle\delta(r_{-+})\rangle$ over the distribution $f(\mathbf{R})$, i.e.

\begin{equation}
\label{eq5}
\langle\delta(r_{-+})\rangle =
\frac{\int f(\mathbf{R}) \delta(r_{-+}) d\mathbf{R}}
{\int f(\mathbf{R}) d\mathbf{R}}
\end{equation}

\noindent
Recalling that $f(\mathbf{R})$ is symmetric under any exchange between
electrons, Eq. \ref{eq5} can rewritten as

\begin{equation}
\label{eq6}
\langle\delta(r_{-+})\rangle =
\frac{\int \rho(\mathbf{r}_-,\mathbf{r}_+) \delta(r_{-+}) d\mathbf{r}_-d\mathbf{r}_+}
{\int \rho(\mathbf{r}_-,\mathbf{r}_+) d\mathbf{r}_-d\mathbf{r}_+}
\end{equation}
\noindent
where $\rho(\mathbf{r}_-,\mathbf{r}_+) = N_{ele} \int f(\mathbf{R}) 
d\mathbf{r}_2 ...d\mathbf{r}_N$. Introducing the new coordinates 
$\mathbf{R}_{-+} = \mathbf{r}_- + \mathbf{r}_+$ and
$\mathbf{r}_{-+} = \mathbf{r}_+ - \mathbf{r}_-$, after integration over
$\mathbf{R}_{-+}$ and spherically averaging over $\mathbf{r}_{-+}$ one gets

\begin{equation}
\label{eq7}
\langle\delta(r_{-+})\rangle =
\frac{\int \Omega(\mathbf{r}_{-+}) \delta(r_{-+}) d\mathbf{r}_{-+}}
{\int \Omega(\mathbf{r}_{-+}) d\mathbf{r}_{-+}} =
\frac{\int \Omega(r_{-+}) \delta(r_{-+})r_{-+}^2 d{r}_{-+}}
{\int \Omega(r_{-+}) r_{-+}^2dr_{-+}} =
\frac{\Omega(0)}
{\int \Omega(r_{-+}) r_{-+}^2dr_{-+}}
\end{equation}
\noindent
where $\Omega(r_{-+})$ is the spherically averaged positron-electron
distribution.
Although these manipulations are quite straightforward, they highlight
that in order to compute $\langle\delta(r_{-+})\rangle$ one must
have accurate values for both $\Omega(0)$ and the denominator
$\int \Omega(r_{-+}) r_{-+}^2dr_{-+}$. Therefore, both the coalescence
region and the tail of the distribution must be correctly described.

In order to thoroughly present the complexity of the faced problem,
Figure 1 shows a typical behavior of $\Omega(r_{-+})$ as sampled
from the model wave function for one electron and one positron

\begin{equation}
\label{eq8}
\Psi_1(\mathbf{r}_{-},\mathbf{r}_{+}) =
\exp[ -r_- - 0.25r_+ -0.25r_{-+}]
\end{equation}
\noindent
by means of a standard VMC simulation using the Langevin algorithm
and the accept/reject step ~\cite{reybook}. This
simulation was carried out sampling a grand total of $3.75\times10^9$ 
configurations and using a time step of 0.01 hartree$^{-1}$, 
a fairly small time step for this simple wave function. 
The sampled distribution of $r_{-+}=r$ was collected on a grid with a
bin width of $\delta r=$0.025 bohr, and then the number
of times $r_{-+}$ was found inside a given bin was divided by
the volume of the spherical crown $[r-\delta r$,$r]$, 
$V(r,\delta r)= \frac{4\pi}{3}[3r^2\delta r -3r\delta r^2 +\delta r^3]$.
The so obtained values were attributed to the mean radius of the spherical 
crown, $\overline{r}=\pi\delta r [4r^3 -6r^2\delta r+4r \delta r^2
-\delta r^3]/V(r,\delta r)$. This is equivalent to approximating
$\Omega(r)$ as a straight line inside each spherical crown, a fairly good 
approximation in such a small bin.

We want to stress that the shape of the distribution
was found independent of the time step over a broad range of values. This
ensures that no systematic bias is present due to the finite time step.
From Figure 1 one can easily notice the abrupt decrease of the distribution
in the region close to $r=0$. This represents the aforementioned
inability of Monte Carlo simulations to correctly sample the distribution
close to a coalescence point in spite of the large number of sampled 
configurations. It also seems to indicate that, due to this inability, 
any well-based method (e.g. both the spherical well and Gaussian methods)
should return an inaccurate answer for r$_w$ smaller that a 
certain threshold.
Conversely, all the regions with $r>0.5$ bohr seem to be adequately described
by the sampled distribution, and therefore we propose to analytically continue
their shape extrapolating to $r=0$ by means of a suitable
functional form. This idea allows one to exploit the knowledge
about the exact form of $\Omega$ to improve the local
description in the small radius regions. For instance, if one samples
$f=\Psi_0^2$, the exact value of the cusp condition can be used as a way
to constrain the model to behave correctly. This trick can also be used
in both the VMC and DMC simulations, since it is often easy to
obtain the cusp condition of the sampled $f$ knowing the analytical form
of $\Psi_T$.
This method could be implemented in two different ways. First, one could
choose an analytical function $\omega(r)$ to fit $\Omega$ for all the 
electron-positron distances. This function should be flexible enough to
properly describe both short range and long range behavior of $\Omega$.
More specifically, close to $r=0$ $\Omega(r)$ behaves like 
$\exp(-\alpha r)$, where $\alpha$
is strictly related to the cusp condition. Differently, in the large r
regions $\Omega(r)$ follows $\exp(-\beta r)$, where $\beta$ is dependent
on the positron affinity (PA) of the system.
A possible choice for $\omega(r)$ is the Pad\'e-Jastrow form

\begin{equation}
\label{eq9}
\omega(r) = N_{\omega}\exp[-\frac{\alpha r + \beta r^2}{1+\gamma r}]
\end{equation}
\noindent
where $\alpha$ can be chosen in order to have $\omega(r)$ satisfying the
correct local behavior close to $r=0$.
The fitted form can successively be used to estimate both $\Omega(0)$ and the 
denominator in Eq. \ref{eq7}.

Second, if the form of $\Omega(r)$ is more complicated (e.g. it has 
multiple maxima), it is possible to resort
to a local fit by $\omega(r)$ in the region close to the cusp in order to get the
$\Omega(0)$ value. Then, the normalization integral could be split in 
two part, one computed using $\omega(r)$ and the other directly using the sampled
distribution. Specifically for the distribution in Figure 1, 
one could fit the sampled $\Omega$ values
in the range [0.5,1.0] bohr constraining $\omega(r)$ to have both the exact cusp
behavior and to have the same value of the sampled $\Omega$ for r=1.0 bohr.
Then, the normalization integral can be estimated integrating
numerically $\omega(r)$ for $0 \leq r \leq 1.0$, and employing the trapezoidal
formula for the remaining sampled values. We would like to mention that
this necessity is already present for small systems like e$^+$Be 
~\cite{maxmgbe}.

Although the two proposed methods are  approximate, 
they might prove themselves to be quite accurate in practice,
allowing the Monte Carlo practitioner to easily estimate the collision
probability between two particles. In turn, this will allow to compute the
annihilation rate in positronic systems, therefore opening the chance 
to directly compare with the experimental results.

In order to show that this is exactly the case, in the next section we 
present the results obtained computing $\langle \delta(r_{-+}) \rangle$
for some model systems whose exact values are easily
obtained using different methods.

Methods similar to ours, although quite different in many details, have been 
applied by Ortiz ~\cite{ortiz} and by Mair\'i Fraser ~\cite{fraser} 
to the case of a positron embedded in
the jellium. Their methods were somehow tailored to the specific system
under study, so that no direct comparison with our proposals can be made.
Nevertheless, the results they extracted from the simulations can be useful
to seek for possible correlations between the magnitude of the "pile up"
effect and the local electron density.

\heading{III. TEST OF THE METHODS USING MODEL SYSTEMS}

To test the accuracy of the proposed methods, we computed the 
$\langle \delta(r_{-+}) \rangle$ expectation value for simple model systems
composed only by one electron and one positron. More specifically,
some model wave functions $\Psi_{i}$ were chosen in order to represent the
variety of electron, positron and electron-positron distributions
that could be found in a positron atomic system.
Then, the $\Psi_{i}^2$'s were sampled by means of VMC simulations
similar to the one discussed above, in order to collect the electron-positron
distribution $\Omega$. 

We selected three model systems as representative of
a fairly large class of positron complexes.
The first one is given by the wave function $\Psi_1$ (see Eq. \ref{eq8}).
The second has the analytical wave function

\begin{equation}
\label{eq10}
\Psi_{2}(\mathbf{r}_{-},\mathbf{r}_{+}) =
\exp[ -r_- + \frac{0.15r_+ -0.5r_+^2}{1+r_+} -0.5r_{-+}]
\end{equation}
\noindent
where the simple exponential in $r_+$ of $\Psi_{1}$ is substituted
by a Pad\'e-Jastrow and the exact cusp condition between electron and
positron has been introduced.

To mimic the presence of core and valence shells, we chose
as a third function

\begin{equation}
\label{eq11}
\Psi_{3}(\mathbf{r}_{-},\mathbf{r}_{+}) = 
\{\exp[-\frac{r_- +2r_-^2}{1+r_-}] + 0.001 \exp[\frac{15r_- -3r_-^2}{1+r_-}]\}
\exp[ \frac{0.15r_+ -0.5r_+^2}{1+r_+} -0.5r_{-+}]
\end{equation}
\noindent

\noindent
To compute the exact value $\langle \delta(r_{-+}) \rangle$ for these
models we used its definition

\begin{equation}
\label{eq12}
\langle \delta(r_{-+}) \rangle = \frac{\int \Psi_{i}^2
(\mathbf{r}_-,\mathbf{r}_+) \delta(r_{-+}) 
d\mathbf{r}_+d\mathbf{r}_-}{\int \Psi_{i}^2(\mathbf{r}_-,\mathbf{r}_+)
d\mathbf{r}_+d\mathbf{r}_-}=
\frac{\int \Psi_{i}^2 (\mathbf{r}_-,\mathbf{r}_-)
d\mathbf{r}_-}{\int \Psi_{i}^2(\mathbf{r}_-,\mathbf{r}_+)d\mathbf{r}_+
d\mathbf{r}_-}
\end{equation}
\noindent
The simple radial integral at the numerator was computed
by numerical integration on a grid, while the denominator was estimated by
means of Eq. \ref{eq3} using $g(\mathbf{r}_-,\mathbf{r}_+) =
A^3B^3 \exp[-2Ar_+ -2Br_-]/\pi^2$.

The two distributions sampled as a function of $r$ from $\Psi_{2}$ and 
$\Psi_{3}$ turned out to possess a behavior quite similar to $\Psi_{1}$. 
Therefore, we avoid to show all of them and refer to Figure 1 as a template
for such distributions. Due to their smoothness,
we fitted them using Eq. \ref{eq9} over the range 0.3-10 bohr
constraining the Pad\'e-Jastrow form to have the exact cusp condition,
i.e. -0.5, -1.0, and -1.0, for the three $\Psi_i$ respectively.
To test the correctness of the chosen fitting range, we modified slightly
the lower limit without finding statistically meaningful differences.
Then, the fitted $\omega$ was used to estimate both $\Omega(0)$ and the
denominator in Eq. \ref{eq7} by means of numerical integration.
The computed results, shown as $\langle\delta(r_{-+})\rangle_{Pade'}$,  
are presented in Table \ref{tab1} together with the "exact" values
computed using Eq. \ref{eq12}.

During the simulations
we also computed the mean values $\langle G(r_{-+},\gamma) \rangle$,
where $G(r_{-+},\gamma)=N_{\gamma} \exp[-r^2_{-+}/\gamma]$ is a 
normalized Gaussian function, for $\gamma =$ 0.01, 0.0033, 0.002, and 0.001.
This technique was proposed by Kenny {\it et al.} ~\cite{needs} 
and successively applied in Ref. ~\cite{morosi} and Ref. ~\cite{maxcor}.
The $\langle G(r_{-+},\gamma) \rangle$ 
values were extrapolated to $\gamma=0$ by fitting them with the 
simple function $a\sqrt{\gamma} +b$, the extrapolation
law deduced in Ref. ~\cite{maxcor} using model systems. 
The fitting was quite accurate for all the three cases, and the
results for $b=\langle G(r_{-+},0) \rangle$ are also shown in Table \ref{tab1}.

Comparing $\langle\delta(r_{-+})\rangle_{Pade'}$ with the exact results
it strikes the really good agreement between these two sets of values,
the relative error being 1\% at most for all the models. 
It must be pointed out that this level of relative accuracy is 
sufficient to thoroughly compare with the experimental data.
It is also worth to stress that the application to other model systems 
gave a similar or better
relative accuracy, therefore showing the wide applicability of the method.

As already pointed out previously ~\cite{mecor,morosi}, also the extrapolated 
$\langle G(r_{-+},0) \rangle$ values are in good agreement with the exact 
results.  Nevertheless, one should expect to obtain really inaccurate
approximations to the exact results using $\langle G(r_{-+},\gamma) \rangle$
with $\gamma$ smaller than some threshold value. 
This simple idea is based on the incorrect sampling of the density
close to $r=0$ shown in Fig. 1, so the good agreement found in this and 
previous works calls for an explanation.
This is easily obtained superimposing $G(r_{-+},\gamma)r_{-+}^2$ to the
sampled $\Omega(r_{-+})$, i.e. comparing the behavior of the two
factors that form the function whose integrals must be estimated.
It turns out that $G(r_{-+},\gamma)r_{-+}^2$ for $\gamma \geq0.001$
has its largest values where $\Omega(r_{-+})$ still behaves correctly, therefore
allowing a correct estimate of the integrals. Tests carried out using 
smaller values of $\gamma$ gave much worse results than the ones reported, 
so we believe it is safer to limit the values of this parameter to the 
range 0.001-0.01 in order to obtain a meaningful extrapolation. 
Although this may look some way problematic due to the aforementioned 
difficulties, from the results in Table \ref{tab1} $\langle G(r_{-+},0) \rangle$
appears to be a good first estimate of the exact $\langle\delta(r_{-+})\rangle$.
In conclusion, we suggest Monte Carlo practitioners to
carry out always both estimations, i.e. extrapolating
$\langle G(r_{-+},\gamma) \rangle$ and fitting the sampled $\Omega$,
as a way to safely estimate $\langle \delta(r_{-+}) \rangle$.

As far as diffusion Monte Carlo and the exact sampling of $\Psi^2_0$
are concerned, the application of these methods
is straightforward, and no more complications are expected than in the VMC case.

\heading{IV. THE e$^+$LiH SYSTEM}

Having verified the accuracy of the proposed method in computing 
Dirac's delta mean values, we applied it to the calculation
of the annihilation rate $\Gamma_{2\gamma}$ of e$^+$LiH for various internuclear
distances R. 

Although this system has already been carefully studied employing
both QMC methods ~\cite{maxcor,maxe+lih} and 
Explicitly Correlated Gaussian (ECG) functions ~\cite{stras3,stras4,mitroylih},
a description of $\Gamma_{2\gamma}$ as a function
of the molecular geometry is still lacking. Up to now, there are only 
$\langle\delta(r_{-+})\rangle=\langle\delta(r_{1+})\rangle+
\langle\delta(r_{2+})\rangle+\langle\delta(r_{3+})\rangle+
\langle\delta(r_{4+})\rangle$ results at R=3.015 bohr 
(0.0240(8) from DMC simulations ~\cite{maxcor}
and 0.024992 form ECG calculations ~\cite{stras3}), at the estimated 
equilibrium distance R=3.348 bohr
(0.027252) ~\cite{stras4}, and the non-adiabatic results of Mitroy and Ryhzikh,
0.034016 and 0.032588 ~\cite{mitroylih}. 
These last values were obtained using ECG in connection , respectively, with
the Stochastic Variational Minimization (SVM) and the Frozen-Core SVM (FCSVM) 
methods, and are roughly 15-20\% larger
than the ECG ~\cite{stras3,stras4} and DMC ~\cite{maxcor} clamped-nuclei ones.
This unexpected result lead Strasburger ~\cite{stras3} to consider the 
possibility of the flattening of the potential
energy curve of e$^+$LiH with respect to the LiH one, a feature that
may allow the positronic molecule to visit the large internuclear distance 
region where the Dirac's delta mean value is expected to be larger. 
However, Mitroy and Ryhzikh ~\cite{mitroylih} pointed out that simple ECG's
may not represent the best basis functions to describe vibrational nuclear 
motion, and that their vibrational averaged nuclear distances are probably
too large.

In a previous work ~\cite{maxe+lih}, we computed the complete curve using the 
DMC technique,
showing that the flattening is indeed present, and that a strong red shift 
of the vibrational spectrum with respect to LiH must be expected.
Unfortunately, due to the high computational
cost of our highly correlated trial wave functions, at that time we did 
not compute the behavior of 
$\langle \delta \rangle$ as a function of R.
In this work we still adopt the BO approximation, so we
predict the annihilation rate for each vibrational state of the system
and compare our values with the non-adiabatic results of Mitroy
and Ryhzikh ~\cite{mitroylih}.

In order to study the effect of the molecular geometry on the annihilation
rate without using the computationally expensive wave function used
in Ref. ~\cite{maxcor}, we decided to employ a model potential approach
to eliminate the ``core'' electrons of the Li$^+$ fragment. We believe
such an approach to be physically well grounded, as explained by the
following supporting reasons.
First, the ECG calculations carried out by Strasburger ~\cite{stras3,stras4} on
the e$^+$LiH system show that the annihilation takes place primarily
with the two electrons that may be attributed to the H$^-$ fragment.
Secondly, the frozen core approximation developed by Mitroy and Ryhzikh
~\cite{newjim} has been found to describe accurately the annihilation process 
in e$^+$Li, e$^+$Be, LiPs, and e$^+$He $^3$S when compared
with the corresponding all-electron calculations.

In order to reduce the number of active electrons, we used for e$^+$LiH the 
model Hamiltonian

\begin{equation}
\label{eq13}
\mathcal{H}_{mod} = -\frac{1}{2}[\nabla^2_1 + \nabla^2_2 + \nabla^2_+]
+V^e_{mod}(\mathbf{r}_1)+V^e_{mod}(\mathbf{r}_2)-\frac{1}{r_{H1}}
-\frac{1}{r_{H2}}
+\frac{1}{r_{12}} +\frac{1}{r_{H+}}-\frac{1}{r_{1+}}-\frac{1}{r_{2+}}
+V^+_{mod}(\mathbf{r}_+)
\end{equation}
Here, the $r_{ij}$ are interparticle distances, 1 and 2 being the electrons,
+ the positron, and $H$ the hydrogen nucleus. Moreover, the 
Bardsley's model potential ~\cite{bard}

\begin{equation}
\label{eq14}
V^e_{mod}(\mathbf{r}_i)=\frac{-1+10\exp[-2.202r_{i\mathrm{Li}}]}
{r_{i\mathrm{Li}}}
\end{equation}
\noindent
where $r_{i\mathrm{Li}}$ is the distance between the $i$-th electron and the
Li nucleus, has been used to represent the 1s$^2$ Li$^+$ core electrons.
To model the interaction of the positron with the frozen Li$^+$ fragment
we simply added to the repulsive Coulomb potential of the nucleus
the potential of the two frozen core electrons as described by an STO-1s orbital
with Z=3, obtaining

\begin{equation}
\label{eq15}
V_{mod}^+(\mathbf{r}_+)=\frac{3}{r_{\mathrm{L}i+}} 
+6\exp[-6r_{\mathrm{Li}+}]
-2\frac{1-\exp[-6r_{\mathrm{Li}+}]}{r_{\mathrm{Li}+}}
\end{equation}
where $r_{\mathrm{Li}+}$ is the Li-positron distance.

To test the accuracy of this model potential, we computed the energy for the
ground state of the three systems Li, Li$^-$, and LiPs. The energy values
are respectively -0.1953(2) hartree, -0.2191(3) hartree, and
-0.4485(3) hartree. They give an electron affinity of 0.0238(4) hartree,
a positron affinity (PA) of 0.2294(4)
hartree, and a Ps binding energy (BE) of 0.0032(4) hartree.
While the electron affinity turns out to be in fair agreement
with the experimental value, namely 0.023 hartree ~\cite{expea}, 
both PA and BE are roughly
0.009 hartree smaller than the best estimate ~\cite{newjim}.
These discrepancies might be due to the absence of polarization effects
of the core electrons due to the two active electrons and the positron,
and to a relatively inaccurate representation of the 1s$^2$ electron
density by means of a single STO-1s function.
Nevertheless, since we are primarily interested in obtaining a semiquantitative
description of the changes in $\Gamma_{2\gamma}$ for this system, we
believe the approximations introduced in the model Hamiltonian to
be small enough to allow for a correct prediction of the trend for this
important observable.

In order to accurately describe the wave function of the
three active particles at the VMC level, we employed a trial wave function
form similar to the one used in Ref. ~\cite{maxcor}, but slightly
modified to include the polarization of the
electron and positron density of the PsH fragment due to the Li$^+$
potential. Specifically, the analytical form in Eq. 7 of 
Ref. ~\cite{maxcor} has been multiplied by a Pad\'e-Jastrow factor depending
on the $z$ coordinate of each particle: the $z$ axis was chosen as the LiH bond
axis, the H nucleus being located at the origin and the Li on the negative $z$
axis. The wave function parameters
were fully optimized for every nuclear distance R minimizing the
variance of the local energy over a fixed sample of configurations
~\cite{frost,conroy}.
This procedure is already well described in the literature ~\cite{reybook}, so
we skip the unnecessary details.
The ensemble of walkers used in the optimization was generated
by DMC simulations in order to bias the walkers
distribution towards the exact density. Usually, four or five
optimization steps were carried out for each R. 

We started the optimization process of the wave function
at R=20 ~\cite{maxcor}. 
At the end of the optimization procedure, R was decreased 
and the wave function re-optimized for the new distance. 
This procedure gives the
chance to monitor the changes of the wave function with R, 
but might increase the possibility to remain stuck in a local minimum 
in the parameter space during the optimization. 

Having optimized at VMC level the approximate wave functions for various 
distances,
these were employed in long DMC simulations to project out the remaining
excited state contributions and to compute more accurate mixed expectation
values. For all the simulations, a time step of 0.005 hartree$^{-1}$
was used, together with a population of 9000 walkers. These two simulation
parameters were found adequate to make statistically negligible both the 
time step bias and the population effect in the DMC simulations.

The DMC results for the energy and for the $\langle \delta(r_{-+}) \rangle$
of this model system are shown in Table \ref{tab2}. There, the energy values 
represent the ground state energy of the model Hamiltonian Eq. \ref{eq13},
$\langle \delta(r_{-+}) \rangle_{Pade'}$ are the total collision probabilities
estimated using the electron-positron distribution, i.e. 
$\langle \delta(r_{1+}) \rangle +\langle \delta(r_{2+}) \rangle$,
while $\langle G(r_{-+},0) \rangle$ are the extrapolated Gaussian
values. Here, the electron-positron distributions were fitted with
the function in Eq. \ref{eq9}, constraining its cusp to be -0.5+cusp$(\Psi_T)$.

The energy values obtained in Ref. ~\cite{maxe+lih}, 
after having subtracted the repulsion 1/R between the
H nucleus and the Li$^+$ core and the
total energy of the Li$^+$ fragment (-7.279913 hartree ~\cite{mitroy})
to estimate the leptonic energy of the PsH moiety, are shown in
Figure 2 together with the DMC results obtained in this work.

For R$\geq$4, the results from the model system follow closely
the more accurate all-electron FN-DMC values, showing that the model potential
correctly describes the polarization of PsH due to the interaction with Li$^+$.
For shorter distances the approximation of considering Li$^+$ frozen is
no longer accurate, so that a discrepancy between the two sets of results
is expected.
It is also important to remember that the Bardsley's potential
was tailored only to describe the atom in its ground and
valence excited states, not to describe correctly bonds in molecules. 

Figure 3 shows the two computed $\langle \delta(r_{-+}) \rangle$ values
as a function of the internuclear distance R, together with the ECG results of
Strasburger ~\cite{stras3,stras4} and the DMC result of Ref. ~\cite{maxcor}.
These last results can be used to evaluate the total accuracy of our
computed collision probabilities.

From Fig. 3 it is clear that both the 
$\langle\delta(r_{-+})\rangle_{Pade'}$ and the 
$\langle G(r_{-+},0)\rangle$ are in good agreement, the second
differing from the first one by two standard deviations at most.
This allows us to think that we are accurately estimating
the mixed distribution mean value obtained by the standard DMC technique. 
Improved results could be obtained only by sampling $\Psi_0^2$.

As far as the total accuracy is concerned, at R=3.015 bohr
both all-electron ECG ~\cite{stras3,stras4} and DMC results
~\cite{maxcor} appear to be smaller than the model ones by roughly 7\%. 
Instead, on going towards large R the mean value seems to correctly converge 
to the very accurate
ECG value, namely $\langle \delta(r_{-+}) \rangle_{PsH}=$0.04874 ~\cite{mitroy},
and to the new DMC 0.0486 estimate carried out in this
work using the electron-positron distribution.
From this comparison, one could expect our estimation to slightly degrade
on going towards small R, without becoming embarrassing 
inaccurate to create concerns about the usefulness of this model system.

The overall trend of the collision probability shows a net decrease
on going towards short R, a feature already suggested by Mitroy and Ryzhikh
~\cite{mitroylih}.
This can be easily understood remembering that
the Li$^+$ model potential repels the positron, while attracting the two
electrons. It is not easy to infer any possible analytical model
to describe these joined effects, although for large R
one could propose a limiting
1/R$^2$ form due to the polarization of the two distributions
by the electric field of Li$^+$. We show in the Appendix that this
reasoning is indeed correct for any observable by means of first order 
Perturbation Theory. An ``experimental'' evidence that this is the case
is given by the fairly good fitting of the $\langle \delta \rangle$ results at 
R=10, 15, and 20 bohr with the simple
form $0.0486+b/R^2$, where b = -0.20448.

Various other mean values were computed during the DMC simulations in 
order to obtain some physical insight on the electron and positron density
behavior. Figure 4 reports the mean value of the $z$ coordinate for the 
two particles giving information on the polarization of the two
lepton densities. It is clear that the
positronic distribution is polarized by the model potential in the opposite
direction to the electronic one. Moreover, it appears to be more easily
polarized than the electronic one always showing larger $\langle z \rangle$
values. This fact can be easily explained by noticing that the positron
distribution is more diffuse that the electron one, so that it is
more strongly repelled by the electric field. 
Interestingly, at R=2 and 2.5 bohr the electron distribution 
reverse its
polarization showing $\langle z \rangle>0$. We believe this effect to be due
to the repulsive region of the Li$^+$ core potential that pushes away
the electrons for such small nuclear distances, therefore modeling 
the exchange effect created by an antisymmetric wave function.

Figure 5 shows the average values for the electron-electron and 
electron-positron distances.  While the
electron-positron mean distance $\langle r_{-+} \rangle$ increases 
monotonically upon decreasing R, the electron-electron distance 
$\langle r_{--} \rangle$ shows
a shallow maximum around R=8 bohr and a deep minimum around R=2.5 bohr.
We believe the maximum to be due to the competition between the positron 
and the Li$^+$ model potential to bind an electron. More specifically,
although polarized towards positive $z$, the positron still attracts one of the 
two electrons to form the Ps sub-cluster, increasing the distance from the 
second electron that is free 
to be polarized in the direction of the Li$^+$ core.
On going towards smaller R, the positron is pushed far out
the bond region, loosing its ability to polarize the electrons that are now
both strongly attracted by the model potential. This interaction leads them
to move in the small volume between H and Li$^+$,
therefore decreasing their mean distance.
Then, for R smaller than 2.5 bohr, the electron-core 
repulsion pushes the electrons outwards the bond 
region, with the net effect to increase their mean distance.
This effect has also been observed by plotting the
intracule electron distributions obtained during the DMC simulations.

It is worth to mention that similar conclusions can be drawn analysing the
VMC results obtained as a by-product of the optimization stages.

Having studied the overall behavior of $\langle \delta \rangle$, 
$\langle z \rangle$, and $\langle r \rangle$, we now
turn to compute the vibrationally averaged annihilation probabilities.
To obtain these quantities, we interpolated our
$\langle \delta \rangle$ results by means of the analytical form
$\Delta(R)=0.0486 -2aR/(1+bR+cR^2+dR^3)$.
The fitted parameters are $a=1.05945$, $b=97.5779$, $c=-37.9705$, and
$d=12.2715$.
Then, the potential energy curve of e$^+$LiH obtained in Ref. ~\cite{maxe+lih}
was fitted with the modified Morse potential 

\begin{equation}
V_M(R)=
-8.0699+A\{1-\exp[-B(R-C)]\}^2-A
-D\{1-\exp[-(R/F)]^6\}/(2R^4)
\label{eq15bis}
\end{equation}
\noindent
obtaining $A=0.03444$ hartree, $B=0.72030$ bohr$^{-1}$, $C=3.3060$ bohr,
$D=21.1796$ bohr$^{-3}$, and $F=5.88217$ bohr.
The last term in Eq. \ref{eq15bis} has been introduced in order to correctly
represent the charge-induced dipole interaction between Li$^+$ and PsH.
The nuclear Schr\"odinger equation for this potential was then solved
using the grid method proposed by Tobin and Hinze ~\cite{tobin}, 
and the numerical
wave functions $\phi_{\nu}(R)$ were then used to compute vibrationally 
averaged mean values for zero total angular momentum. 
More specifically, we computed

\begin{equation}
\label{eq16}
\langle O \rangle_{\nu} = 
\frac{\int d\mathbf{R} \phi^2_{\nu}(R) O(R)}{\int d\mathbf{R} \phi^2_{\nu}(R)}
\end{equation}
\noindent
where $O(R)$ is $\Delta(R)$ or any other function of $R$.

In Table \ref{tab3} we show the results for $\langle \delta \rangle_{\nu}$
and $\langle R \rangle_{\nu}$
computed over the first 16 bound vibrational states. 
The $\langle \delta \rangle_{\nu}$values 
increase in an almost linear fashion on going towards large $\nu$, 
as expected by the steady increase
of $\langle R \rangle_{\nu}$ due to the vibrational excitation.
Comparing $\langle \delta \rangle_{0}=$0.0295 with the value of $\Delta(R)$ at 
the equilibrium distance of our fitted potential, namely 0.0291 at 3.353 bohr, 
it appears that the ground level vibrational motion only slightly increases the
probability of collision between the electrons and the positron with respect
to the one at the equilibrium distance. This finding is in line with the small
difference between the equilibrium distance and the average nuclear distance
$R_0=$3.42 bohr.  We relate these outcomes to the almost linear behavior of
$\langle \Delta(R) \rangle$, and to the shape of 
$R^2\phi_0^2(R)$ in the region around the potential minimum, where it resembles
a Gaussian.

Although our vibrationally averaged result for $\nu=0$ 
$\langle \delta \rangle_0$= 0.0295 appears to be roughly 
8\% larger than the ECG result (0.027252) at the equilibrium distance 3.348 bohr
~\cite{stras4} suggesting a fairly large effect of the nuclear motion,
we believe this outcome is primarily due to the 7\% larger collision 
probabilities computed using our model system.
These two evidences seem to rule out the Strasburger's suggestion ~\cite{stras3}
of a large increase of the collision probability
due to the quantum nuclear motion for the $\nu=0$ state. They indicate that
approximating the averaged collision probability for the vibrational 
ground state simply by using its
value at the equilibrium distance could be a fairly accurate procedure.
Moreover, these conclusions agree with Mitroy and Ryhzikh ~\cite{mitroylih} 
warnings that both the SVM and FCSVM results,
although proving the overall stability of e$^+$LiH,
are not well converged to the exact ones. For instance, their
$\langle R \rangle_0$ values, respectively 4.182 bohr and 
3.964 bohr, are larger than the minimum of the ECG and DMC
potential curves by more than 0.5 bohr. 
This discrepancy cannot be accounted for by the zero point 
motion of the positron complex.
These larger distances between the two fragments Li$^+$ and PsH
in the non-adiabatic treatment imply a reduced distortion of the lepton 
densities of PsH with respect to the one at the Born-Oppenheimer equilibrium,
and therefore a too large annihilation rate.
However, it is interesting to notice that both FCSVM (0.032588) and SVM 
(0.034016) ~\cite{mitroylih} collision probabilities are really close to our 
Born-Oppenheimer one at $R=$4.0 bohr. In our view, this agreement stresses,
again, the importance of the local electric field in defining the collision
probability and the overall accuracy of the SVM approach in describing
the relative densities an a positronic complex.

As far as the behavior of $\langle \delta(R) \rangle_{\nu}$ is concerned,
the steady increase on going towards large $\nu$ indicates that the annihilation
rate does depend on the quantum vibrational state of the molecule.
Although the trend of these results could be specific of the e$^+$LiH 
system and perhaps of other polar molecules as well, 
it strongly indicates that any theory formulated
to describe ``on the fly'' annihilation of e$^+$ due to Feshbach resonances
must include this effect in order to go beyond ``order of magnitude
comparison'' ~\cite{gribakin} and to predict accurately the annihilation rate.
In our view, this opens a new avenue of exploration in positron physical 
chemistry where the understanding of the vibrational motion effect on
positron annihilation by molecular systems is of prime importance.

\heading{V. CONCLUSIONS}

In this work we have critically compared methods that may be useful
to compute the annihilation rate in positronic systems in the framework of 
the QMC methods.
Moreover, we have presented a simple, but nevertheless solid and accurate,
method based only on the interparticle distribution sampling.
After having tested it using model systems, we employed the method
to compute  $\langle \delta(r_{-+}) \rangle$ for e$^+$LiH
for several internuclear distances. These results allowed us to
discuss many interesting features of this positronic complex, and to
predict that the annihilation probability increases upon increasing
the vibrational quantum number $\nu$. We notice that a
similar behavior of $\langle \delta(r_{-+}) \rangle$ may be expected
also for e$^+$LiF due to the polarization of the positronic density
of the PsF fragment by the Li$^+$ core. The situation could be quite different
for the e$^+$BeO case where the positron density is expected to be centered on
Be at large nuclear distances (the two fragments e$^+$Be and O have lower total
energy than Be$^+$ and PsO ~\cite{maxmole}), and to move on the O side of the 
molecule when the distance decreases. 
This effect is due to the electron transfer from Be to O that 
creates the large molecular dipole moment. From our experience on these systems
we expect  $\langle \delta(r_{-+}) \rangle$ for e$^+$Be to be smaller than
the one for the polar molecule, so the vibrationally 
excited states close to the dissociation threshold may have smaller 
annihilation rates than the ground vibrational state. 

It is also interesting to speculate on the behavior of the annihilation rate
versus the vibrational quantum number for other simple systems like
e$^+$Li$_2$ and e$^+$Be$_2$. Here, the symmetry of the systems can play an
important role in defining the annihilation rate. For instance, decreasing
the nuclear distance one may expect to find the positron localized between the
two atomic fragments due to its ability to polarize the two atomic electron
densities: in this situation the annihilation rate could be quite different
from the atomic one. Although it is easy to infer the existence of a
bound state for these complexes at large nuclear distances  employing the basic 
Valence Bond resonance idea

\begin{equation}
\mathrm{e^+A}_1 +\mathrm{A}_2 \leftarrow\rightarrow \mathrm{A}_1 
+ \mathrm{e^+A}_2
\label{eq17bis}
\end{equation}
\noindent
it still remains to demonstrate the 
stability of these systems for nuclear distances close to the equilibrium 
geometry of the neutral parent molecules.

As rule of thumb to predict the stability of a non-polar molecule,
one can use the adiabatic ionization potential (AIP) as proposed 
by Mitroy {\it et al.} ~\cite{modelalk}. 
For the X $^1 \Sigma _g ^+$ ground state of Li$_2$
the experimental AIP is 0.189 hartree ~\cite{expli2}, 
slightly lower than the atomic one, 0.19814 hartree ~\cite{explibe}. 
Also for the X $^1 \Sigma_g^+$ ground state of Be$_2$,
one might expect a similar lowering of the AIP with respect
to the atomic one, 0.343 hartree ~\cite{explibe}, 
so that a value of around 0.335 hartree could be regarded as a
safe upper bound to the true AIP. Both these values fall inside 
the upper and lower IP limits for positron binding obtained by Mitroy 
{\it et al.} ~\cite{modelalk} for one and two valence electron atoms, 
therefore suggesting that the two complexes should be stable. 
We understand that this model
is just a rough approximation for our molecular systems ~\cite{maxansw}.
Nevertheless, a positron bound to an atom or a molecule is always characterized 
by a quite diffuse density. This allows one to neglect some of the real features
of the electron density close to the nuclei as a first approximation, and 
focus only on the asymptotic properties of the positron cloud that are
correlated to the IP and to the polarizability.

As far as Be$_2$ is concerned,
the AIP larger than the Ps binding energy (0.25 hartree) suggests a mechanism 
based on the electron cloud polarization as responsible of the binding. 
Moreover, the lowering of the AIP with respect to the atomic one, and the large
polarizability of this molecule (roughly twice the atomic one) seem to indicate
its ability to form a stronger bond with the positron than the Be atom alone
~\cite{newjim}.
Conversely, Li$_2$ has an AIP smaller than the Ps binding energy,
suggesting that the polarization of the Ps cluster may be held responsible
for the positron binding. However, Li is close to the lower stability threshold
of the positron-atom complexes, and we do not feel confident in proposing
the stability of the molecular complex with respect to the
Ps+Li$_2^+$ dissociation pathway.

Right now, QMC methods are the best suited computational techniques to carry 
out such a study since 6 and 8 electron systems are too large to be studied
with ECG's unless the frozen core approximation is used. 
With the addition to the QMC ``bag of tricks'' of a robust method for 
computing annihilation rates such a study could become routine in
molecular physical chemistry, allowing the exploration of many interesting 
features of these ``exotic'' compounds.

Moreover, many more other technically oriented applications could be devised.
Positronium annihilation in polymers and membranes, positron annihilation
in silicon nanocluster, nanodevices, 
fullerenes and carbon nanotubes are just few of them that could be 
quite easily interpreted with the help of such a method.

Our hope is that this work will help this kind of applications to blossom
and to lead to a better understanding of the basic interaction schemes
that positron has with ordinary matter.

\heading{ACKNOWLEDGMENTS}
MM acknowledges Jim Mitroy for many interesting discussions on positronic
systems, as well as for his help in improving the manuscript.
Financial support by the Universita' degli Studi di Milano 
is gratefully acknowledged. 
The authors are indebted to the Centro CNR per lo Studio delle Relazioni
tra Struttura e Reattivita' Chimica for grants of computer time.

\pagebreak
\heading{APPENDIX}

In this Appendix we show that the correction to the expectation value of 
every observable $O$ for PsH interacting 
with the Li$^+$ core follows the limiting analytical form
$R^{-2}$ for large $R$. This is indeed a general results for positronic
atom systems immersed in a weak electric field. 
From Perturbation Theory one can write the first order corrected wave function
for the ground state as

\begin{equation}
\Psi_0^{(1)} = \Psi_0^{(0)} + \sum_{i\not=0} \frac{\int\Psi_i^{(0)} V
\Psi_0^{(0)} d\mathbf{R}}{E_0^{(0)}-E_i^{(0)}}\Psi_i^{(0)} 
= \Psi_0^{(0)} + \sum_{i\not=0} c_i^{(0)} \Psi_i^{(0)} \nonumber
\label{eq17}
\end{equation}
\noindent
where $\Psi_i^{(0)}$ are the eigenstates of the unperturbed Hamiltonian
(i.e. the PsH), $E_i^{(0)}$ its eigenvalues, and $V$ the perturbation 
potential. The expectation value of the observable $O$ can be computed
using

\begin{equation}
\langle O \rangle = \frac{\mathbf{c}^t\mathbf{O}\mathbf{c}}
{\mathbf{c}^t\mathbf{I}\mathbf{c}} \nonumber
\label{eq18}
\end{equation}
\noindent
where $\mathbf{O}_{ji} = \int \Psi_j^{(0)} O \Psi_i^{(0)} d\mathbf{R}$,
while $\mathbf{c}=\{1,c_1^{(0)},c_2^{(0)},...\}$.
If the perturbation potential is small with respect to
the total energy, then  $\mathbf{c}^{(0)}_i \ll 1$ and 
$\mathbf{c}^t\mathbf{I}\mathbf{c} =1+ \sum_{i\not=0}
(\mathbf{c}^{(0)}_i)^2 \simeq 1$. This fact allows to introduce a further
approximation into Eq. \ref{eq18},

\begin{equation}
\langle O \rangle \simeq \mathbf{c}^t\mathbf{O}\mathbf{c} =
\langle O \rangle_{00} +2\sum_{i\not=0}\mathbf{c}^{(0)}_i
\langle O \rangle_{0i}+\sum_{i,j\not=0}\mathbf{c}^{(0)}_j\mathbf{c}^{(0)}_i
\langle O \rangle_{ji} \simeq
\langle O \rangle_{00} +2\sum_{i\not=0}\mathbf{c}^{(0)}_i
\langle O \rangle_{0i}
\label{eq18bis}
\end{equation}
\noindent
showing that the first order change in the expectation value $\langle O \rangle$
is linearly dependent on the $\mathbf{c}^{(0)}_i$'s.

For our specific case, namely PsH interacting with the Coulomb potential
of Li$^+$ at distance $R$, for $R\rightarrow \infty$ the perturbing 
interaction potential can be written as

\begin{equation}
V = \sum_k \frac{q_k}{R_{q_kLi}} \simeq \sum_k q_k(\frac{1}{R} - 
\frac{z_k}{R^2})
\label{eq19}
\end{equation}
\noindent
where the molecular geometry is as in the main text, while $q_k$ are the 
leptonic charges.
This approximation is equal to consider the electron and positron densities
constant in a plane parallel to the $xy$ plane. Introducing this approximation
in the integrals in Eq. \ref{eq17} one gets

\begin{eqnarray}
\int\Psi_i^{(0)} V \Psi_0^{(0)} d\mathbf{R} \simeq
\sum_k q_k\frac{1}{R}\int\Psi_i^{(0)} \Psi_0^{(0)} d\mathbf{R} +
\sum_k q_k \int\Psi_i^{(0)} \frac{z_k}{R^2} \Psi_0^{(0)} d\mathbf{R} = \\ 
\nonumber
\frac{\sum_k q_k\int\Psi_i^{(0)} z_k \Psi_0^{(0)} d\mathbf{R}}{R^2}
\label{eq20}
\end{eqnarray}

This result shows that the $\mathbf{c}_i$ in Eq. \ref{eq18} and
\ref{eq18bis} are proportional
to $1/R^2$, therefore proving that this is also the analytical
form of the leading correction to the unperturbed ground state expectation 
values $\mathbf{O}_{00}$.

\clearpage

\heading{Figure Captions:}

Figure 1: Behavior of $\Omega(r_{-+})$ sampled from $\Psi_{1}$.

Figure 2: Energy of the PsH moiety computed from Eq. 13 and
from the e$^+$LiH results of Ref. \cite{maxe+lih} as explained in the text.

Figure 3: Computed $\langle \delta \rangle$ results for e$^+$LiH
system at various internuclear distances.

Figure 4: Electron and positron mean value of the $z$ coordinate as a function
of the internuclear distance.

Figure 5: Electron-electron and electron-positron mean distances as 
function of the internuclear distance.

\clearpage

\begin{table}
\begin{center}

\begin{tabular}{llccc}  \hline\hline
 &$\langle\delta(r_{-+})\rangle_{exact}$&$\langle\delta(r_{-+})\rangle_{Pade'}$&$\gamma$&$\langle G(r_{-+},\gamma) \rangle$\\ \hline
 $\Psi_{1}$   &    && 0.0100   & 0.02140(8)  \\
              &    && 0.0033   & 0.0219(1)  \\
              &    && 0.0020   & 0.0222(2)  \\
              &    && 0.0010   & 0.0223(3)  \\
              & 0.022602(7)   &0.02229 & 0.0000   & 0.0228(3)   \\ \hline
 $\Psi_{2}$   &    && 0.0100   & 0.0979(2)  \\
              &    && 0.0033   & 0.1028(3)  \\
              &    && 0.0020   & 0.1043(4)  \\
              &    && 0.0010   & 0.1058(7)  \\
              & 0.10981(1)   &0.11052& 0.0000   & 0.1095(7)   \\ \hline
 $\Psi_{3}$   &    && 0.0100   & 0.0842(2)  \\
              &    && 0.0033   & 0.0886(3)  \\
              &    && 0.0020   & 0.0901(4)  \\
              &    && 0.0010   & 0.0916(7)  \\
              & 0.09399(1) & 0.09382 & 0.0000   & 0.0950(7)  \\
\hline \hline

\end{tabular}
\caption{$\langle \delta(r_{-+}) \rangle$ expectation values for the
three model systems $\Psi_{1}$, $\Psi_{2}$, and $\Psi_{3}$. 
The "exact" values are computed by Eq. \ref{eq12}. 
$\langle\delta(r_{-+})\rangle_{Pade'}$ are computed fitting Eq. \ref{eq9}
to the sampled distribution as explained in the text.
$\gamma$ is the width of the Gaussian used to compute 
$\langle G(r_{-+},\gamma) \rangle$.}
\label{tab1}
\end{center}
\end{table}

\clearpage

\begin{table}
\begin{center}

\begin{tabular}{lccc}  \hline\hline
R &$\langle E\rangle$&$\langle\delta(r_{-+})\rangle_{Pade'}$&$\langle G(r_{-+},0) \rangle$\\ \hline
2.0 & -1.16437(9)   & 0.01884  & 0.0182(2)  \\
2.5 & -1.15936(7)   & 0.02178  & 0.0210(2)  \\
3.0 & -1.13196(6)   & 0.02684  & 0.0256(2)  \\
3.5 & -1.09798(5)   & 0.03000  & 0.0288(2)  \\
4.0 & -1.06485(5)   & 0.03334  & 0.0324(2)  \\
6.0 & -0.96985(4)   & 0.04166  & 0.0402(2)  \\
8.0 & -0.91995(3)   & 0.04544  & 0.0436(3)  \\
10.0 &-0.89160(3)   & 0.04666  & 0.0448(3)  \\
15.0 &-0.85634(2)   & 0.04754  & 0.0458(3)  \\
20.0 &-0.83932(2)   & 0.04794  & 0.0468(3)  \\
$\infty$& -0.78918(1)& 0.04860 & 0.0484(3)  \\
\hline \hline

\end{tabular}
\caption{ Lepton energies, $\langle\delta(r_{-+})\rangle_{Pade'}$, and
$\langle G(r_{-+},0) \rangle$ mean values for the e$^+$LiH model
system. All quantities in atomic units.}
\label{tab2}
\end{center}
\end{table}

\clearpage

\begin{table}
\begin{center}

\begin{tabular}{lcc}  \hline\hline
$\nu$  &$\langle\delta \rangle_{\nu}$\\ \hline
 0  &  0.0295 & 3.423 \\
 1  &  0.0305 & 3.571 \\
 2  &  0.0315 & 3.732 \\
 3  &  0.0325 & 3.906 \\
 4  &  0.0334 & 4.094 \\
 5  &  0.0344 & 4.294 \\
 6  &  0.0352 & 4.503 \\
 7  &  0.0361 & 4.718 \\
 8  &  0.0369 & 4.938 \\
 9  &  0.0376 & 5.166 \\
 10 &  0.0383 & 5.408 \\
 11 &  0.0390 & 5.675 \\
 12 &  0.0397 & 5.985 \\
 13 &  0.0405 & 6.356 \\
 14 &  0.0414 & 6.803 \\
 15 &  0.0421 & 7.243 \\
\hline \hline
\end{tabular}
\caption{Vibrationally averaged $\langle\delta(r_{-+})\rangle_{\nu}$ and
$\langle R\rangle_{\nu}$
 values for the e$^+$LiH model system. All quantities in atomic units.}
\label{tab3}
\end{center}
\end{table}


\clearpage

\begin{thebibliography}{99}

\bibitem{moge} O. E. Mogensen, {\it Positron annihilation in chemistry}, Springer-Verlag, Berlin (1995).
\bibitem{kraus} R. Krause-Rehberg and H. S. Leipner, {\it Positron annihilation in semiconductors}, Springer-Verlag, Berlin (1999).
\bibitem{mitroy} G. G. Ryzhikh, J. Mitroy, and K. Varga, J. Phys. B {\bf 31}, 3965 (1998).
\bibitem{newjim} J. Mitroy and G. G. Ryzhikh, J. Phys. B  {\bf 34}, 2001 (2001).
\bibitem{jiang_psh} N. Jiang and D. M. Schrader, J. Chem. Phys. {\bf 109}, 9430 (1998).
\bibitem{stras3} K. Strasburger, J. Chem. Phys. {\bf 111}, 10555 (1999).
\bibitem{stras4} K. Strasburger, J. Chem. Phys. {\bf 114}, 615 (2001).
\bibitem{mecor} D. Bressanini, M. Mella, and G. Morosi, Phys. Rev. A {\bf 55}, 200 (1997).
\bibitem{morosi} D. Bressanini, M. Mella, and G. Morosi, Chem. Phys. Lett. {\bf 272}, 370 (1997).
\bibitem{maxpsh} D. Bressanini, M. Mella, and G. Morosi, Phys. Rev. A {\bf 57}, 1678 (1998).
\bibitem{maxcor} M. Mella, G. Morosi, and D. Bressanini, J. Chem. Phys. {\bf 111}, 108 (1999).
\bibitem{brom1} M. W. J. Bromley, J. Mitroy, and G. G. Ryzhikh, Nucl. Inst. and Meth. B {\bf 171}, 47 (2000).
\bibitem{reybook} B. L. Hammond, W. A. Lester, Jr., and P. J. Reynolds, {\it Monte Carlo Methods in Ab Initio Quantum Chemistry}, 1st ed., World Scientific, Singapore (1994).
\bibitem{langf} P. Langfelder, S. M. Rothstein, and J. Vrbik, J. Chem. Phys. {\bf 107}, 8525 (1997).
\bibitem{caffarel} R. Assaraf and M. Caffarel, Phys. Rev. Lett {\bf 83}, 4682 (1999).
\bibitem{caffarel2} R. Assaraf and M. Caffarel,  J. Chem. Phys. {\bf 113}, 4028 (2000).
\bibitem{coldw} S. A. Alexander and R. L. Coldwell, J. Chem. Phys. {\bf 103}, 2572 (1995).
\bibitem{spiega1} The ``pile up'' effect has been found by us to account at least for half of the annihilation probability of a positron bound to a molecular system.
\bibitem{spiega2} Our earlier attempts to introduce explicit correlation between the electrons and the positron did not obtain any success. We believe this outcome to be primarly due to a lack of flexibility of the trial wave function that was not able to account for the relevant changes in the electronic distributions after the positron addition. Moreover, the minimization of the variance of the local energy resulted in a reduction of the positron-nuclear repulsion instead of increasing the correlation between electrons and the positron.
\bibitem{and75} J. B. Anderson, J. Chem. Phys. {\bf 63}, 1499 (1975).
\bibitem{and76} J. B. Anderson, J. Chem. Phys. {\bf 65}, 4121 (1976).
\bibitem{rey82} P. J. Reynolds, D. M. Ceperley, B. J. Alder, and W. A. Lester, Jr., J. Chem. Phys. {\bf 77}, 5593 (1982).
\bibitem{pshpol} M. Mella, D. Bressanini, and G. Morosi, Phys. Rev. A {\bf 63}, 024503 (2001).
\bibitem{reytag} R. N. Barnett, P. J. Reynolds, and W. A. Lester, Jr., J. Comput. Phys. {\bf 96}, 258 (1991).
\bibitem{rept} S. Baroni and S. Moroni, Phys. Rev. Lett. {\bf 82}, 4745 (1999).
\bibitem{umri_dmc} C. J. Umrigar, M. P. Nightingale, and K. J. Runge, J. Chem. Phys. {\bf 99}, 2865 (1993).
\bibitem{maxmgbe} M. Mella, (Unpublished). The electron-positron distributions in the e$^+$Be and e$^+$Mg systems show minima and maxima that are fingerprints of the multi-shell atomic structure. The same effect is much less evident for LiPs.
\bibitem{ortiz} G. Ortiz, PhD thesis (Ecole Polytechnique Federale de Lausanne, 1992).
\bibitem{fraser} M. Mair\'i Fraser, PhD thesis (Imperial College of London, 1997).
\bibitem{needs} S. D. Kenny, G. Rajagopal, and R. J. Needs, Phys. Rev. A {\bf 51}, 1898 (1995).
\bibitem{maxe+lih} M. Mella, G. Morosi, D. Bressanini, and S. Elli, J. Chem. Phys. {\bf 113}, 6154 (2000).
\bibitem{mitroylih} J. Mitroy and G. G. Ryzhikh, J. Phys. B {\bf 33}, 3495 (2000).
\bibitem{bard} J. N. Bardsley, "Pseudopotentials in atomic and molecular physics", Case Studies in Atomic Physics {\bf 4}, 299 (1974).
\bibitem{expea} H. Hotop and W. C. Lineberg, J. Phys. Chem. Ref. Data {\bf 14}, 731 (1985).
\bibitem{frost} A. A. Frost, J. Chem. Phys. {\bf 41}, 478 (1964).
\bibitem{conroy} H. Conroy, J. Chem. Phys. {\bf 41}, 1327 (1964).
\bibitem{tobin} F. L. Tobin and J. Hinze, J. Chem. Phys. {\bf 63}, 1034 (1975).
\bibitem{gribakin} G. F. Gribakin, Phys. Rev. A {\bf 61}, 022720 (2000).
\bibitem{maxmole} D. Bressanini, M. Mella, and G. Morosi, J. Chem. Phys. {\bf 109}, 1716 (1998).
\bibitem{modelalk} J. Mitroy, M. W. J. Bromley, and G. Ryzhikh, J. Phys. B {\bf 32}, 2203 (1999).
\bibitem{expli2} R. A. Bernheim, L. P. Gold, and T. Tipton, J. Chem. Phys. {\bf 78}, 3635 (1983).
\bibitem{explibe} G. L. Sharon, I. E. Bartmess, J. F. Liebman, J. L. Holmes, R. D. Levin, and W. G. Hallard, J. Phys. Chem. Ref. Data {\bf 17}, Suppl. 1 (1988).
\bibitem{maxansw} M.Mella, G. Morosi, and D. Bressanini, J. Chem. Phys. {\bf 112}, 3928 (2000).

\end{thebibliography}
\end{document}